# KubeEdge-Sedna v0.3：
# 迈向次时代自动定制的AI工程范式

## KubeEdge-Sedna v0.3: Towards Next-Generation Automatically Customized AI Engineering Scheme

华为云边缘云创新实验室 郑子木

摘要：全球边缘AI市场规模持续增长。当前阻碍边缘规模复制的技术挑战主要是边侧小样本与边缘数据异构。另外，边缘AI客户往往存在对数据安全合规与边缘AI服务离线自治需求。基于学界终身学习方法，我们首次正式地定义边云协同终身学习问题，发布业界首个开源边云协同终身学习。边云协同终身学习通过（1）多任务迁移学习适应不同边缘局点数据异构，实现"千人千面"精准预测；（2）未知任务增量处理在小样本下越学越聪明，逐步实现AI工程化与自动化；（3）借助云侧知识库来记忆新情景知识，避免灾难性遗忘；（4）边云协同架构使得在应用云上资源的同时保证数据安全合规与边缘AI服务离线自治，希望从根本上解决上述边云协同机器学习的挑战。

关键词：边缘AI；边云协同；终身学习

## 1 当前机器学习落地技术挑战

### 1.1 当前机器学习落地有哪些问题？

近二十年来，机器学习已广泛应用于数据挖掘、计算机视觉、自然语言处理、生物特征识别、搜索引擎、医学诊断、检测信用卡欺诈、证券市场分析、DNA序列测序、语音和手写识别、战略游戏和机器人等领域。

在实际业务落地过程中，大部分大型云平台提供商均已提供机器学习算力等资源服务，同时支持多种机器学习框架等以提供开放灵活的部署环境。但是，机器学习模型所需的数据往往并非从云平台中产生，而是从传感器、手机、网关等边缘设备中产生。数据从边侧产生，而云端需从边侧采集数据以训练和不断完善机器学习模型。在实际落地时，当前机器学习需面对以下问题：

（1）海量设备数据导致延迟和成本问题

· 即使有100Mbps的专网连接，将10TB的数据运送到云端也需要10天。

· 大量边缘连接设备每天生成数百兆字节甚至TB数据，带来的延迟和成本对客户和服务提供方来说往往是难以承受的。

（2）数据压缩导致的延迟和精度问题

· 迁移所有数据通常不切实际，往往需要对数据进行"压缩"（如特征工程、难例识别等）并传输到云端，而数据压缩过程容易引入新的延迟。

· 压缩数据不一定能代表完整数据集信息，容易导致精度损失。

（3）边缘数据隐私和计算实时性问题

上述问题的本质来源是数据在边缘产生，而算力却在云端更为充足。也就是说，在机器学习服务将边缘产生的数据转换为知识的过程中，一方面需要在边缘快速响应并处理本地产生的数据，另一方面需要云上算力与开发环境的支持。随着边缘设备数量指数级增长以及设备性能的提升，边云协同机器学习应运而生，以期打通机器学习的最后一公里。

### 1.2 当前边云协同机器学习落地有哪些技术挑战

目前边云协同机器学习的经典模式是：在云上给定一个数据集，运行机器学习算法构建一个模型，然后将这个模型不作更改应用在多个边侧的多次推理任务上。这种学习范式称为封闭学习（也称孤立学习[1]），因为它并未考虑其他情景学习到的知识和以往学习到的知识。虽然边云协同机器学习技术的相关研究和应用都有着显著的进展，然而在成本、性能、安全方面仍有诸多挑战：数据孤岛/小样本/数据异构/资源受限[2]。

在边缘云背景下，不同边缘数据分布总是不断变化，而边缘标注样本也往往由于成本较高而数量稀少。因而封闭学习需不停标注样本并重新训练，这显然给服务落地带来巨大挑战。这种数据分布和数据量上的挑战分配称为数据异构和小样本，属于边云协同机器学习的四大挑战之一。

本文以一个热舒适预测服务示例介绍相应挑战，如图1所示。该服务输入外界温度等环境特征，预测不同人员的热舒适程度（热、舒适、冷）。由于边缘节点部署位置从室外变动到室内，对于相同室外温度特征值x=30，可以看到实际标注的热舒适标注发生了较大变动。原有室外模型上线预测值整体偏低，要匹配到室内模型，则需要训练样本重新调整。即面对分布动态变化

的不同边侧数据，由于没有记忆历史和不同情景任务知识，封闭学习需要频繁重新训练。

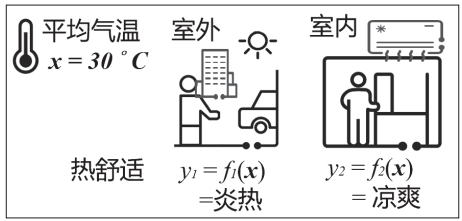

图1 热舒适预测服务中机器学习模型随边侧
环境变化示意图

### 1.3 当前边云协同机器学习技术挑战应如何解决？

从上面的讨论中可以了解到，当前的封闭学习范式可被用于提供数据同构和大数据的服务，但难以处理数据异构和小样本的问题，所以并不合适用于建立通用的机器学习系统。伊利诺伊大学芝加哥分校的刘兵教授也在Frontiers of Computer Science中总结，封闭学习范式一系列局限性的根本在于没有记忆，这导致它通常需要大量的训练样本。

对应的范式改进可以从人类的学习过程中得到启发。可以看到，人类之所以能够越学越聪明，是由于每个人并非自我封闭地学习，而是不断地积累过去学习的知识，并利用其他人的知识，学习更多知识[1]。借鉴人类这种学习机制，提出终身学习结合边云协同可以发展出边云协同终身学习。边云协同终身学习通过多任务迁移学习适应不同边缘节点数据异构，实现"千人千面"精准预测；未知任务增量处理在小样本下越学越聪明，逐步实现AI工程化与自动化；借助云侧知识库来记忆新情景知识，避免灾难性遗忘；边云协同架构使得在应用云上资源的同时保证数据安全合规与边缘AI服务离线自治，从根本上解决上述边云协同机器学习的挑战。

## 2 边云协同终身学习概念

基于1995年学界提出的终身学习概念[3]，进一步定义边云协同终身学习为边云协同的多机器学习任务持续学习。其中机器学习任务是指在特定情境下运用的模型，如中译英（给定汉语翻译成英语）、亚洲植物分类等。正式定义如下：

边云协同终身学习：给定云侧知识库中N个历史训练任务，推理持续到来的当前任务和未来M个边侧任务，并持续更新云侧知识库。其中，M趋向于无穷大，同时边侧M个推理任务不一定在云侧知识库N个历史训练任务当中。

具体来说，边云协同终身学习的一般流程如图2所示。

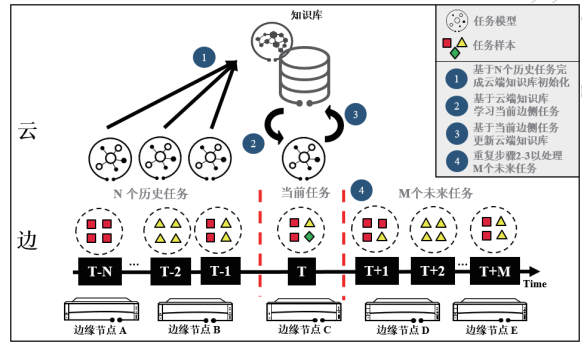

图2 边云协同终身学习流程示意图

（1）初始化知识库：在云侧知识库中存储和维护过去N个任务（记为第T-N到T-1个任务）中训练并累积的知识。

（2）学习当前任务：在边侧设备面对当前任务（记为第T个任务）时，基于云侧知识库先验知识训练第T个任务。注意，第T个任务并不一定在历史的N个任务当中。

（3）更新知识库：将学习到的边侧第T个任务知识反馈给云侧知识库并更新。

（4）学习未来任务：持续学习未来M个任务（记为第T+1到T+M个任务）。与上面第T个任务利用过去N个任务知识（从T-N到T-1）类似，第T+1个任务的边侧任务知识则利用过去N+1个云侧知识（从T-N到T）。以此类推，直到完成第T+M个任务，结束整个流程。

边云协同终身学习具备以下三大功能点：

（1）边云协同持续学习：能够基于云侧算力和边侧数据合作完成持续推理与训练，能够在推理运行时增强模型训练能力。

（2）边云协同多任务知识维护：以云侧知识库作为中心，实现跨边云的任务知识共享，处理边侧任务同时维护云端知识。

（3）边侧处理云侧未知任务：需要边侧能够发现和处理云端知识库未知任务。其中未知任务是指运行或测试过程中发现的新任务，比如其应用情景或模型已在知识库当前知识之外。

## 3 Sedna边云协同终身学习特性

KubeEdge是一个开源的边缘计算平台，它在Kubernetes原生的容器编排和调度能力之上，扩展实现了边云协同、计算下沉、海量边缘设备管理、边缘自治等能力。KubeEdge还将通过插件的形式支持5G MEC、AI边协同等场景，目前在很多领域都已落地应用[3]。

KubeEdge AI SIG于2020年12月提出了KubeEdge子项目开源平台Sedna，架构如图3所示。Sedna基于

KubeEdge提供的边云协同能力，实现AI的跨边云协同训练和协同推理能力。支持现有AI类应用无缝下沉到边缘，快速实现跨边云的增量学习、联邦学习、协同推理等能力，最终降低边云协同机器学习服务构建与部署成本、提升模型性能、保护数据隐私等[2]。

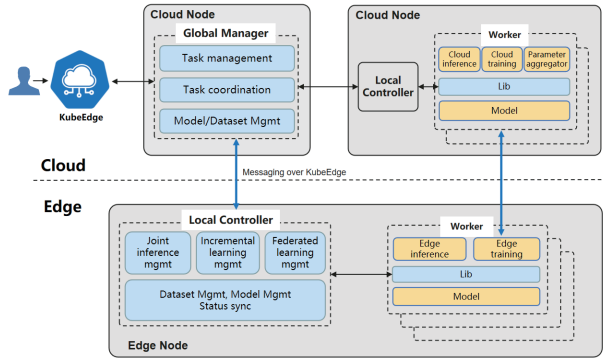

图3 Sedna整体架构

在本次0.3版本更新中，Sedna发布了业界首个边云协同终身学习的开源特性。Sedna终身学习将基于边侧数据和云侧算力，逐步实现适应边侧业务与模型异构"千人千面"的高可信自动化人工智能。

Sedna的边云协同终身学习作业分为三个阶段：训练、评估和部署，维护一个全局可用的知识库（KB）服务于每个终身学习作业（Lifelong Learning Job）。架构如图4所示。

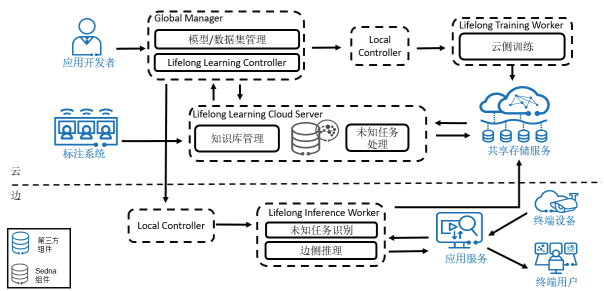

图4 Sedna边云协同终身学习架构

（1）启动Training Worker基于开发者的AI模型和训练数据集进行多任务迁移学习，实现任务的知识归纳，包括：样本属性、AI模型、模型超参参。

（2）Training Worker完成对知识库的更新后，启动评估数据集的Evaluation Worker，基于部署者定义的评估策略判断符合下发部署约束的任务模型。

（3）Global Manager捕获评估任务的完成状态后通知Edge初始化启动Inference Worker进行推理服务。边缘节点基于Sedna Lifelong Learning API进行推理，并进行未知任务上云判别。

（4）通过对接第三方打标系统和知识库的迁移学习，Local Controller基于预配置规则监听新数据变化并按配置的策略触发Training Worker进行增量学习，

重训练完成后重新下发边缘侧。

其中，当前Sedna选用的模块化方案和样本迁移方案使得开源的边云协同终身学习特性能够实现模型无关：（1）同一个特性能够同时支持结构化和非结构化不同模型，在特性中模型可插拔；（2）同一个特性能够同时支持分类、回归、目标检测、异常检测等。

# 4 基于Sedna终身学习实现：楼宇热舒适预测控制

## 4.1 背景

（1）智能楼宇是智慧城市的重要组成部分

楼宇是大量先进工业产品的"使用方"，引领其制造、运行和维护，在这一波能源革命和工业革命中占据重要地位。

现今楼宇都有自控系统，通常它们都在边缘，这使得很多关于楼宇的应用更倾向于部署在边缘侧，其中一类应用是热舒适度预测。由于人们80%的工作和生活都在楼宇中度过，提高工作效率和生活舒适度（如通过楼宇智能化等方式）就显得尤为重要。

（2）热舒适度预测服务于智能楼宇

热舒适度被定义为楼宇中的人对环境冷热的满意程度。它提供了一种定量的评估，把室内冷热环境参数的设定与人的主观评估联系起来。而楼宇中办公或者居住人员的热舒适程度是建筑及其系统设计方案中的一个重要考量因素。在空调系统运作时，一旦热舒适度被预测出来，那么就能将其用于调整楼宇内空调的控制策略。比方说，一种基于热舒适度的控制策略，是基于假定的空调参数设定以及温湿度等环境特征下，给出预计的人体舒适程度，然后搜索寻优出舒适度最高的空调设定。所以，这种情况下要实现舒适度最大的空调控制就依赖于较高精度的舒适度预测。

原有热舒适度的预测要么需要房间中安装额外设备，要么需要人工反馈，部署环境复杂、人工操作频繁使得这种情况下热舒适度的采集准确度非常低。据此，基于机器学习的热舒适度预测方法被提出，它降低部署要求、不需要人工反馈，因而更具有实用价值。

（3）热舒适度预测服务实际部署时数据异构和小样本问题较为突出

由于人员个体差异、房间与城市差异等，不同个体、不同地点对于热舒适的感受是不一样的，那么就会导致相同的环境温度和空调设定下对应的人员的热舒适度标签值不一样，从而导致较为突出的数据异构问题。

热舒适度预测主要针对楼宇中的房间人员个体，具有个性化的特点。在环境因素变化较多的情况下，边侧房间人员个体的热舒适度样本通常有限，往往不足以支

撑对单个人员进行个性化模型的训练，从而导致较为突出的小样本问题。

除了小样本问题之外，增量学习也能够一定程度解决历史与当前情景的数据异构（时间上的数据异构）。但这种边云协同增量学习范式通常不具备用于记忆的知识库，导致很难处理非时间上的数据异构。比方说，对于有多个人员的房间，在同一时刻会存在不同人员上的数据异构。这种情况不仅是同一个人不同时间上的数据异构，增量学习变得不太足够。此时就需要使用边云协同终身学习了。

### 4.2 方案

图5是边云协同终身学习的热舒适预测方案架构，边云协同终身学习的热舒适预测方案主要有两个步骤。

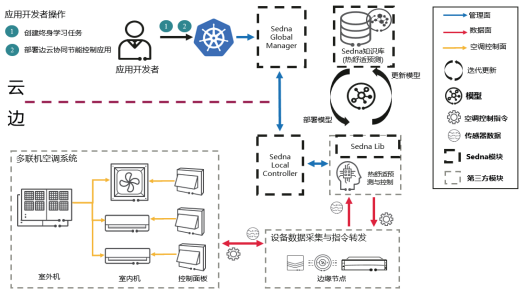

**图5 边云协同终身学习的热舒适预测方案架构**

（1）创建舒适度预测终身学习任务

舒适度预测终身学习任务被创建后，Sedna知识库中会生成舒适度预测的知识库实例，知识库会利用多地点多人员的历史数据集进行初始化，并提供推理和更新接口给边侧应用。

（2）部署边云协同舒适度预测应用

舒适度预测应用被部署后，应用会通过边侧的设备数据采集接口获取到多联机空调系统的设定参数和当前温湿度等环境特征信息。应用通过调用Sedna Lib库终身学习接口，从知识库中寻找对应的任务信息：

·如果被判定为已知任务，例如已经出现过的人员在已知的温湿度条件下，则直接获取对应模型进行推理；

·如果被判定为未知任务，例如是新来的人员，则通过知识库来获取针对未知任务的模型进行推理。并将这些模型和模型之间的关系写入到知识库中，以完成知识库的更新操作，使得知识库得到积累。

### 4.3 效果

本次案例基于开源Ashrae Thermal Comfort II数据集。在这个开源数据集中，收录了全球28个国家99个城市1995~2015年之间楼宇内人员热舒适真实数据，目标是构建一个机器学习分类模型，给定环境特征，预测人群的热倾向（Thermal Preference）。热倾向分为三类，希望更冷（觉得热）、不希望变更（觉得舒适）、希望更热（觉得冷）。

案例结果如图6和图7所示，整体分类精度与单任务增量学习对比，相对提升5.12%（其中多任务提升1.16%）。使用终身学习前后的预测效果在Kota Kinabalu数据中预测率相对提升24.04%，在Athens数据中预测率相对提升13.73%。**AP**

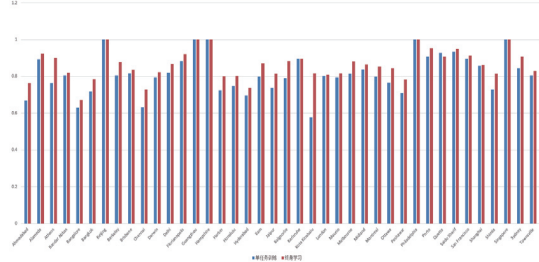

**图6 ATCII各城市Sedna终身学习预测精度对比**

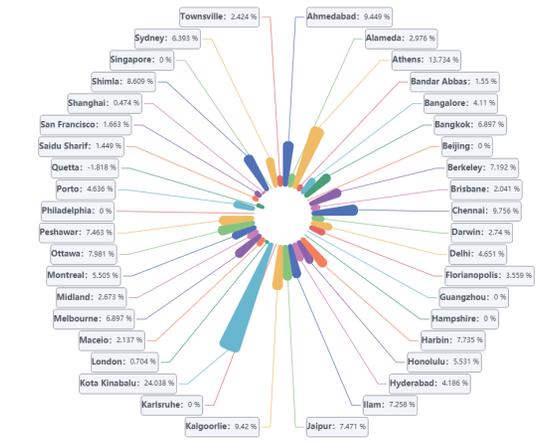

**图7 ATCII各城市Sedna终身学习预测精度相对提升**

---

**作者简介：**

**郑子木**，博士，现就职于华为云边缘云创新实验室，研究方向为边缘AI、多任务迁移学习及AIoT。发表AI算法与系统、分布式计算等领域国际顶级会议及期刊论文十余篇，多次获得最佳会议论文奖项及华为公司技术贡献奖项。目前正带领团队参与云原生计算基金会KubeEdge SIG AI的开源工作。